\documentclass[amsmath,amssymb,aps,prb,twocolumn,floatfix]{revtex4-2}

\usepackage[caption = false]{subfig}
\usepackage[utf8]{inputenc}
\usepackage{graphicx}
\usepackage{standalone}
\usepackage{dcolumn}
\usepackage{bm}
\usepackage{xcolor}
\usepackage{physics}
\usepackage{amsthm}
\usepackage{enumitem}
\usepackage{mathtools}
\usepackage[colorlinks = true,linkcolor = red,citecolor = magenta]{hyperref}
\usepackage[sort&compress]{natbib}

\newcommand{\pcsadd}{Center for Theoretical Physics of Complex Systems, Institute for Basic Science (IBS), Daejeon 34126, Republic of Korea}
\newcommand{\ustadd}{Basic Science Program, Korea University of Science and Technology (UST), Daejeon 34113, Republic of Korea}

\makeatletter
\renewcommand*{\fnum@figure}{{\normalfont\bfseries \figurename~\thefigure}}
\renewcommand*{\@caption@fignum@sep}{\textbf{:}}
\makeatother

\newcommand{\mh}{\mathcal{H}}
\newcommand{\mhfd}{\mh_\mathrm{FD}}
\newcommand{\mhsd}{\mh_\mathrm{SD}}
\newcommand{\mhabf}{\mh_\mathrm{ABF}}

\newcommand{\mbet}{\langle\tau\rangle}
\newcommand{\mbeet}{\langle\tau\rangle_\mathrm{e}}

\begin{document}

\title{Critical-to-Insulator Transitions and Fractality Edges in Perturbed Flatbands}

\author{Sanghoon Lee}
  \affiliation{\pcsadd}
  \affiliation{\ustadd}

\author{Alexei Andreanov}
  \affiliation{\pcsadd}
  \affiliation{\ustadd}

\author{Sergej Flach}
  \affiliation{\pcsadd}
  \affiliation{\ustadd}

\date{\today}

\begin{abstract}
  We study the effect of quasiperiodic perturbations on one-dimensional all-bands-flat lattice models.
  Such networks can be diagonalized by a finite sequence of local unitary transformations parameterized by angles \(\theta_i\).
  Without loss of generality, we focus on the case of two bands with bandgap \(\Delta\).
  Weak perturbations lead to an effective Hamiltonian with both on- and off-diagonal quasiperiodic terms that depend on \(\theta_i\).
  For some angle values, the effective model coincides with the extended Harper model.
  By varying the parameters of the quasiperiodic potentials, we observe localized insulating states
  and an entire parameter range hosting critical states with subdiffusive transport.
  For finite quasiperiodic potential strength, the critical-to-insulating transition becomes energy dependent with what we term fractality edges separating localized from critical states.
\end{abstract}

\maketitle

\section{Introduction}

One of the most fascinating pursuits in recent decades in condensed matter physics has been to understand the impact of various perturbations on single-particle localized states.
It is well known that states can be localized in the presence of a random disorder or quasiperiodic potential~\cite{anderson1958absence, kramer1993localization, aubry1980analyticity};
however, localization can also be achieved in the absence of disorder.
Certain translation invariant tight-binding networks feature dispersionless Bloch energy bands \(E(\mathbf{k})\equiv E\), known as flatbands~\cite{maksymenko2012flatband,leykam2018artificial,rhim2021singular},
independent of the crystal momentum \(\mathbf{k}\) and implying a macroscopic degeneracy at the energy \(E\).
This is a result of destructive interference caused by network geometry or symmetry, implying zero group velocity \(\nabla_{\mathbf{k}}E\) and a localization of particles in the flatband.
For short-range Hamiltonians, flatbands feature compact localized states, or eigenstates trapped in a strictly finite number of sites~\cite{sutherland1986localization,aoki1996hofstadter}.
Experimentally, compact localized states have been observed in a variety of fine-tuned settings~\cite{leykam2018perspective}.
The interest in flatbands is motivated by their extreme sensitivity to perturbations that lift the macroscopic degeneracy and give rise to unusual behavior and a variety of interesting and exotic phases:
flatband ferromagnetism~\cite{tasaki1992ferromagnetism}, frustrated magnetism~\cite{ramirez1994strongly, derzhko2015strongly}, unconventional Anderson localization~\cite{flach2014detangling}, and superconductivity~\cite{cao2018unconventional}.

Flatband systems can be further fine-tuned to flatten all the dispersive bands, resulting in \emph{all-bands-flat} (ABF) networks~\cite{vidal1998aharonov,danieli2021nonlinear}
that are even more sensitive to perturbations and produce interesting ergodicity-breaking phenomena~\cite{kuno2020flat,danieli2020many,danieli2021nonlinear,danieli2021quantum,vakulchyk2021heat}.
ABF networks can be diagonalized by a finite sequence of non-commuting local unitary transformation into the ABF parent network.
Any perturbation of the original ABF network will result in some nontrivial perturbation of the diagonal parent network which can be thus efficiently analyzed.

The effect of single particle perturbations, e.g., disorder, has been studied for flatband lattices~\cite{vidal2001disorder,flach2014detangling,leykam2017localization,chalker2010anderson,bilitewski2018disordered}.
In particular, unconventional localization length scaling~\cite{flach2014detangling}, analytical mobility edges~\cite{bodyfelt2014flatbands}, and reentrant localization~\cite{goda2006inverse,longhi2021anderson} have been found.
The effects of random disorder on ABF networks have recently been systematically investigated, where nonperturbative delocalization transitions were found~\cite{cadez2021metalinsulator}.

In this work, we perform a systematic study of the impact of weak quasiperiodic perturbation on ABF networks in one dimension, closely following the previous work, Ref.~\onlinecite{cadez2021metalinsulator}.
Without loss of generality, we focus on a two-band ABF ladder.
Then we apply a weak (compared to the bandgap) quasiperiodic perturbation.
We use the smallness of the perturbation to project the Hamiltonian onto a single sublattice, thereby deriving a new effective projected Hamiltonian.
By varying the parameters of the perturbation, we find that the entire spectrum of the projected Hamiltonian is either localized or critical, but never extended, using a mapping to the extended Harper model~\cite{avila2017spectral}.

We also find a transition between localized and critical phases---the \emph{critical-to-insulator transition} (CIT)---that depends on the spatial frequency of the quasiperiodic potential.
No metallic states are found for any values of the parameters.
Upon increasing the strength of the quasiperiodic potential, the critical eigenstates are gradually replaced by localized ones via the appearance of an energy-dependent CIT that we dub \emph{fractality edges}.

The paper is organized as follows.
We start by defining and discussing the construction of the all-bands-flat models in Sec.~\ref{sec:model}.
In Sec.~\ref{sec:weak}, we derive an effective model valid in the limit of weak quasiperiodic perturbation and use it to chart the phase diagram, confirmed numerically.
The properties of the full model at finite perturbation strength are investigated in Sec.~\ref{sec:finite}, followed by conclusions in Sec.~\ref{sec:conclusion}.

\section{The model}
\label{sec:model}

We consider one-dimensional (1D) ABF Hamiltonians.
Without loss of generality, we focus on the case of two bands and hopping between the nearest-neighbor unit cells~\cite{maimaiti2019universal}.
Any such 1D Hamiltonian \(\mhabf\) can be constructed from a macroscopically degenerate diagonal parent matrix \(\mhfd\) with onsite energies \(\varepsilon_a\) and \(\varepsilon_b\) on the two sublattices and the bandgap \(\Delta=|\varepsilon_a-\varepsilon_b|\)~\cite{danieli2021nonlinear}.
We refer to the Hamiltonian in this parent basis as \emph{fully detangled} and consider it to be the parent Hamiltonian for a manifold of ABF systems.
Applying a sequence of local unitary transformations \(U_i\)~\cite{flach2014detangling,danieli2021nonlinear} to \(\mhfd\), we necessarily end up with a connected ABF Hamiltonian.
Throughout the paper, ABF implies a connected ABF network.
For nearest-neighbor unit cell hopping, generically, only two non-commuting local unitary transformations, \(U_1\) and \(U_2\), are needed to generate a 1D ABF Hamiltonian.
In the simplest case, these are parameterized by two angles: \(\theta_{1,2}\) for \(U_{1,2}\), respectively, giving real \(U_{1,2}\).
While there are more possible parameters describing complex \(U_{1,2}\), they do not affect the localization properties discussed below, as demonstrated in Appendix~\ref{app:SU2}.
The full unitary transformation reads \(U = U_2\,U_1\).
We refer the reader to Ref.~\onlinecite{cadez2021metalinsulator} for further details on the construction of ABF Hamiltonians.
The symmetry of the system allows us to consider the irreducible region of \(\theta_{1,2}\in [0, \pi/2]\) only.
We note that this construction method has been conjectured to be exhaustive for the generation of higher dimensional short-range ABF networks as well~\cite{danieli2021nonlinear}.

Next, we add a quasiperiodic perturbation to the ABF Hamiltonian with \(\mh = \mhabf + W\).
Here, \(W\) is defined as a direct sum of \(2\times 2\) matrices \(W(n)\) of all \(1\leq n \leq L\), where \(L\) is the number of unit cells:
\begin{gather}
  W = \bigoplus_{n = 1}^{L}W(n), \qquad W(n) =
  \begin{bmatrix}
    W_{1}(n) & 0 \\
    0 & W_{2}(n)
  \end{bmatrix}.
\end{gather}
The generic form of \(W(n)\) is given by two quasiperiodic fields \(W_{1}\) and \(W_{2}\) defined as follows:
\begin{align}
  W_{1}(n) &= \lambda_{1}\cos(2\pi\alpha n), \notag \\
  W_{2}(n) &= \lambda_{2}\cos(2\pi\alpha n + \beta). 
  \label{eq:fields}
\end{align}
Here, the spatial frequency \(\alpha\) is an irrational number, \(\alpha \in \mathbb{R}\setminus\mathbb{Q}\), \(\beta\) is the phase difference between \(W_{2}\) and \(W_{1}\),
and \(\lambda_{1}\) and \(\lambda_{2}\) are the strengths of the quasiperiodic potentials.

We aim to understand how this perturbation affects the transport properties of the ABF models.
The canonical Aubry--Andr\'e model displays a metal--insulator transition with increasing strength of the potential~\cite{aubry1980analyticity}.
On the other hand, three-dimensional ABF Hamiltonians perturbed by weak random disorder display a nonperturbative metal--insulator transition depending on the Hamiltonian parameters.
For finite disorder strength, a reentrant localization transition is observed~\cite{goda2006inverse}.

\section{Weak quasiperiodic potential and effective projected model}
\label{sec:weak}

\begin{figure}
\centering
  \includegraphics[width = 0.8\columnwidth]{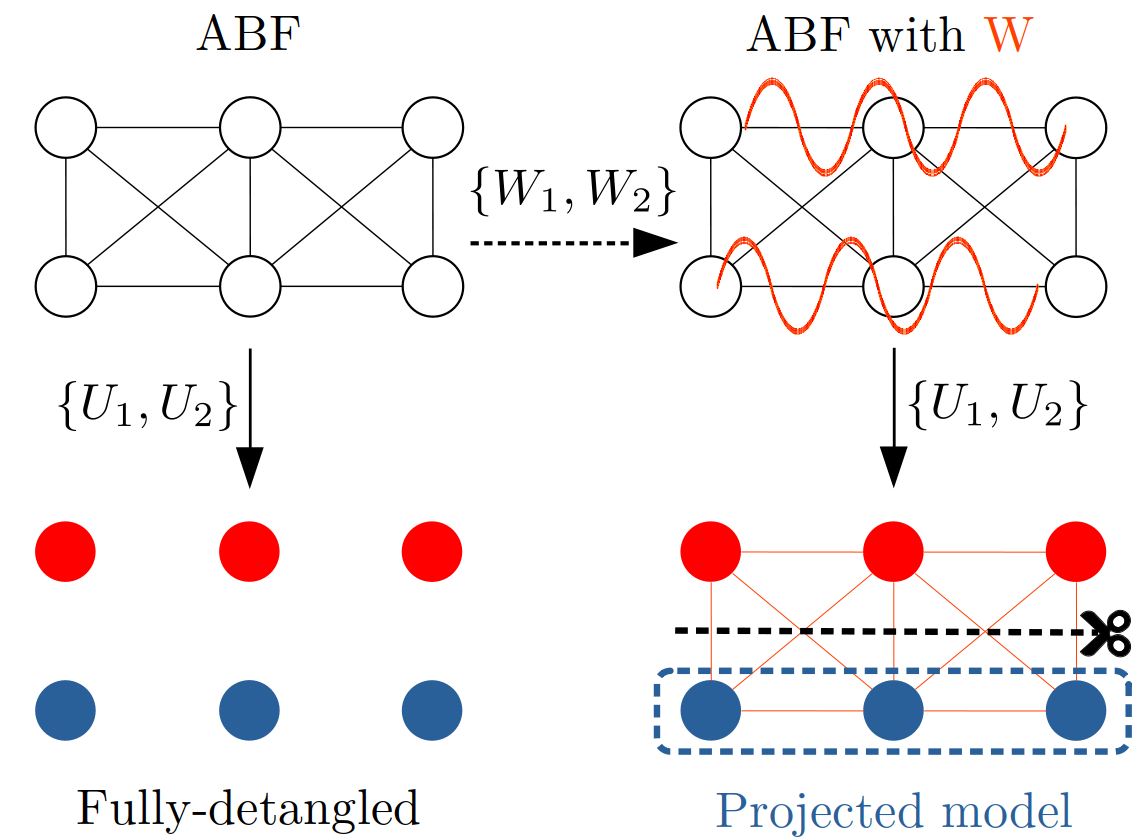}
  \caption{
    Local unitary transformations \(U_{1}\) and \(U_{2}\) convert an ABF model into a fully detangled Hamiltonian (left).
    Addition of quasiperiodic perturbation \(W\) composed of two quasiperiodic fields \(W_{1}\) and \(W_{2}\)~\eqref{eq:fields} creates additional hoppings in the fully detangled basis (right).
    For weak quasiperiodic fields, the first-order degenerate perturbation theory is used to derive the effective projected model.
  }
  \label{fig:scheme}
\end{figure}

We start by examining the limit of weak quasiperiodic perturbation, i.e., vanishing \(\lambda_{1,2}\), as compared to the bandgap \(\Delta\).
This is done by applying the first-order degenerate perturbation theory:
We apply the inverse unitary transformation \(U^\dagger\) to \(\mhabf\), so that the Hamiltonian in a fully detangled basis is \(\tilde{\mh} = \mhfd + U^{\dagger} W U\).
The second term represents hoppings solely due to the quasiperiodic perturbation. The strongest enhancement of the hopping occurs for \(\theta_{1,2} = \pi/4\), which we focus on below since we expect the enhancement to give the strongest delocalizing effect~\cite{cadez2021metalinsulator}.
Without loss of generality, we set \(\lambda_{2} \leq \lambda_{1}\).
Factoring out \(\lambda_{1}\) from \(\tilde{\mh}\) gives us the perturbation \(\tilde{W} = W/\lambda_{1}\), which depends on the ratio \(\lambda_{2}/\lambda_{1} \in [0,1]\) only.
Then via first-order degenerate perturbation theory we get the following eigenvalue problem:
\begin{gather}
  P_{a}U \tilde{W} U^{\dagger}P_{a}\ket{a_{n}} = \lambda_{1}\varepsilon^{(1)}_{a,n}\ket{a_{n}}.
\end{gather}
Here, \(P_{a}\) is a projection operator onto flatband \(\varepsilon_a\) that is local.
The above equation describes an effective 1D tight binding problem that we refer to as the \emph{projected model}~\cite{cadez2021metalinsulator}.
Figure~\ref{fig:scheme} presents the schematics of obtaining the projected model.
On general grounds, we expect the effective model to feature both quasiperiodic onsite energies and finite-range hopping, since \(\tilde{W}\) is local and quasiperiodic, while \(U^\dagger P_a\) is a local operator.
The effective problem for a specific sublattice reads:
\begin{gather}
  \label{eq:pm}
  E a_{n} = v_{n}a_{n} + t_{n}a_{n+1} + t_{n-1}^{*}a_{n-1},
\end{gather}
where the onsite potential \(v_{n}\) and the hopping \(t_{n}\) are indeed quasiperiodic:
\begin{align}
  \label{eq:coefficients:v}
  v_{n} &= v_{s}\sin(2\pi\alpha n - \pi\alpha) + v_{c}\cos(2\pi\alpha n - \pi\alpha), \\
  \label{eq:coefficients:t}
  t_{n} &= t_{s}\sin(2\pi\alpha n) + t_{c}\cos(2\pi\alpha n).
\end{align}
The coefficients \(v_{s,c},t_{s,c}\) depend on the potential strengths \(\lambda_{1,2}\), spatial frequency \(\alpha\), and the phase difference \(\beta\):
\begin{gather*}
  v_{s,c} = v_{s,c}(\lambda_{1,2},\beta,\alpha) \quad\text{and}\quad t_{s,c} = t_{s,c}(\lambda_{1,2},\beta).
\end{gather*}
Their full expressions are provided in Appendix~\ref{app:vt-expr}.
This model describes the transport properties of the ABF network in the weak potential limit, \(\lambda_{1,2}\ll \Delta\).

\subsection{Mapping of the projected model onto the extended Harper model}

The projected model in Eq.~\eqref{eq:pm} features both a quasiperiodic onsite potential and quasiperiodic hopping.
The most generic model of this type is defined by a self-adjoint quasiperiodic Jacobi operator \(J\) acting on a vector of \(l^2 (\mathbb{Z})\) space~\cite{teschl2000jacobi}:
\begin{align}
	(Ju)_{n} &= v(\alpha n + \phi)u_{n} \\
	& + t_{\rho}(\alpha n + \phi)u_{n+1} + t^{*}_{\rho}(\alpha(n-1) + \phi)u_{n-1}, \notag
\end{align}
where \(\alpha\) is an irrational spatial frequency, \(\phi\) is a fixed phase, and \(\rho\) is an additional parameter controlling the hopping \(t\).
For this most generic case, no complete phase diagram or transport properties---numerical or analytical---have been established to the best of our knowledge.
In our case though, the projected model can be mapped onto the already studied extended Harper model for specific parameter values, as discussed below.
The definition of the extended Harper model is given as follows: 
\begin{align}
  \label{eq:EHM}
  (Ju)_{n} &= 2\cos(2\pi\alpha n - \pi\alpha)u_{n} \\
  & + 2\rho\left[\cos(2\pi\alpha n)u_{n+1} + \cos(2\pi\alpha (n-1))u_{n-1}\right]. \notag
\end{align}
Its spectral properties have been characterized completely in Ref.~\onlinecite{avila2017spectral}.
For \(2\rho < 1\), the eigenstates are all localized, as guaranteed by the RAGE theorem~\cite{ruelle1969remark,amrein1973characterization,enss1978asymptotic}.
Otherwise, for \(2\rho \geq 1\), the energy spectrum is fractal~\cite{avila2017spectral}, similar to the Aubry--Andr\'e--Harper model at its critical point.
Moreover, the corresponding eigenstates show multifractal behavior, as demonstrated numerically in~\cite{chang1997multifractal}.

We first cover the choices of parameters \(\lambda_{1,2}, \beta\) that allow our projected model to map onto the extended Harper model.
In these cases, we analytically find critical states and a CIT based on the known phase diagram of the extended Harper model.

\subsubsection{\(\lambda_{2}/\lambda_{1} = 0\)}

When \(\lambda_{2}/\lambda_{1} = 0\), the onsite potential \(v_n\) and the hopping \(t_n\) of the projected model in Eq.~\eqref{eq:pm} are simplified into
\begin{align}
  v_{n} &= \left[\frac{\cos(\pi\alpha)}{4}\right]2\cos(2\pi\alpha n - \pi\alpha),\\
  t_{n} &= \frac{1}{4}\cos(2\pi\alpha n).
\end{align}
The hopping strength \(2\rho\) in Eq.~\eqref{eq:EHM} is always larger than \(1\) regardless of the values of \(\alpha\) and \(\beta\),
\begin{gather}
  2\rho = \left|\frac{1}{\cos(\pi\alpha)}\right| \geq 1,
\end{gather}
implying that all the eigenstates are critical in this case.
Furthermore, since \(\beta\) drops from the effective Hamiltonian, the spectrum and the eigenstates are the same for any value of \(\beta\).
Intuitively, it is clear that if the quasiperiodic potential is only applied to a single leg of the ladder, the phase difference \(\beta\) is irrelevant.

\subsubsection{\(\beta=\pi\) and \(\lambda_{1}\neq \lambda_{2}\)}

When \(\beta=\pi\) and \(\lambda_{1}\neq \lambda_{2}\), the onsite potential and the hopping in Eq.~\eqref{eq:pm} are simplified as follows:
\begin{align*}
  v_{n} &= \left[\frac{1}{4}\left(1-\frac{\lambda_{2}}{\lambda_{1}}\right)\cos(\pi\alpha)\right]2\cos(2\pi\alpha n - \pi\alpha),\\
  t_{n} &= \left[\frac{1}{4}\left(1 + \frac{\lambda_{2}}{\lambda_{1}}\right)\right]\cos(2\pi\alpha n).
\end{align*}
Similarly to the previous case, we find \(2\rho \geq 1\) regardless of the values of \(\alpha\) and \(\lambda_{2}/\lambda_{1}\):
\begin{gather}
  2\rho = \left|\frac{1}{\cos(\pi\alpha)}\right| \left|\frac{1 + \lambda_{2}/\lambda_{1}}{1 - \lambda_{2}/\lambda_{1}}\right| \geq 1.
\end{gather}
This implies that all the eigenstates are always critical.

When the quasiperiodic amplitudes are equal, \(\lambda_1=\lambda_2\), the onsite potential in the projected model [Eq.~\eqref{eq:pm}] vanishes and the strongest hopping can be achieved among other parameter choices.
In this case, the projected model maps onto the self-dual off-diagonal Harper model, which is also critical (see Appendix~\ref{app:self-dual} for details).

\subsubsection{\(\beta = 0\) and \(\lambda_{1}\neq\lambda_{2}\)}
\label{sec:b0-l12}

We now choose \(\beta = 0\) and \(\lambda_{1}\neq\lambda_{2}\). 
The onsite potential and hopping in Eq.~\eqref{eq:pm} are reduced to the following expressions:
\begin{align*}
  v_{n} &= \left[\frac{1}{4}\left(\frac{\lambda_{2}}{\lambda_{1}} + 1\right)\cos(\pi\alpha)\right]2\cos(2\pi\alpha n - \pi\alpha),\\
  t_{n} &= \left[\frac{1}{4}\left(1 - \frac{\lambda_{2}}{\lambda_{1}}\right)\right]\cos(2\pi\alpha n).
\end{align*}
Now \(2\rho\) in Eq.~\eqref{eq:EHM} can be either larger or smaller than \(1\) depending on the values of \(\alpha\) and \(\lambda_{2}/\lambda_{1}\):
\begin{gather}
  \label{eq:transition}
  2\rho = \left|\frac{1}{\cos(\pi\alpha)}\right| \left|\frac{1 - \lambda_{2}/\lambda_{1}}{1 + \lambda_{2}/\lambda_{1}}\right|.
\end{gather}
This leads to a CIT at \(2\rho=1\).
For \(\alpha = (\sqrt{5}-1)/2\), the phase transition point is \((\lambda_{2}/\lambda_{1})_c \approx 0.468\).

We point out that when the quasiperiodic amplitudes are equal (\(\lambda_1=\lambda_2\)) in this case, the hopping \(t_n\) vanishes in the projected model, which makes it diagonal and leaves only the non-zero onsite potential.
Hence, the eigenstates of the projected model exhibit compact localization with eigenstates occupying a single site.
However, since the onsite potential is quasiperiodic, there is no degeneracy in the eigenenergies.
This ultimately leads to the compact localized eigenstates of the original Hamiltonian \(\mh = \mhabf + W\) occupying a finite number of sites without degeneracy.

\subsubsection{\(0<\beta<\pi\) and \(0 < \lambda_2 / \lambda_1 \leq 1\)}
\label{sec:b0p-l12}

\begin{figure}
  \includegraphics[width = \columnwidth]{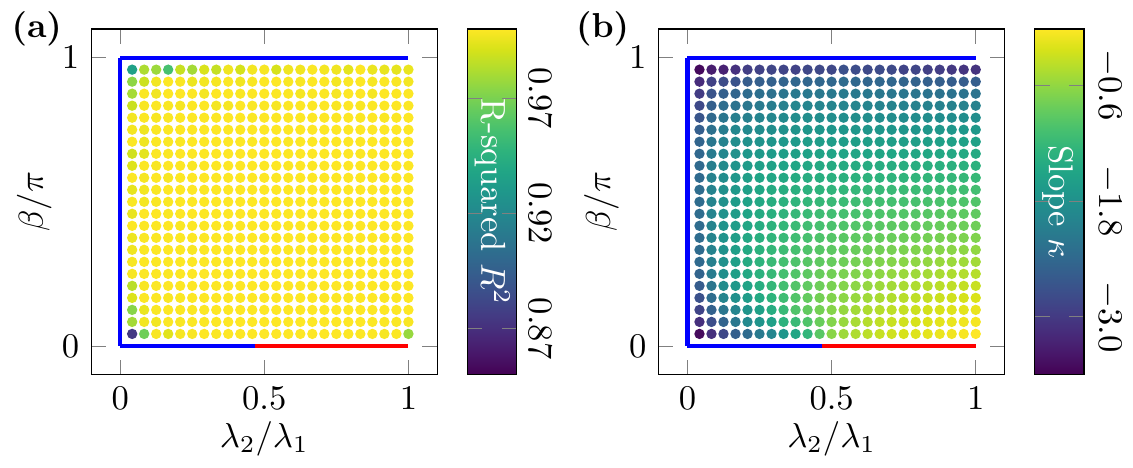}
  \caption{
    Phase diagram of the projected model based on the values of the exponent \(\tau\) averaged over the spectrum, \(\mbet\), computed for \(25\times 25\) values of parameters \(\lambda_2/\lambda_1, \beta/\pi\).
    The blue and red lines on the borders indicate the critical and localized regimes, respectively.
    (a) \(R^2\) values for parameters \(\lambda_2/\lambda_1, \beta/\pi\).
    All points have values very close to \(1\), implying that all eigenstates are localized in the thermodynamic limit in this entire region.
    (b) Slope \(\hat{\kappa}\) for the parameters away from the border of the phase diagram.
    The absolute value of the slope increases closer to the border, implying larger localization length.
  }
  \label{fig:phasediagram}
\end{figure}

We now look into the properties of the states away from the border of the phase diagram shown in Fig.~\ref{fig:phasediagram}, \(0<\beta<\pi\) and \(0 < \lambda_2 / \lambda_1 \leq 1\).
In this case, mapping onto the extended Harper model is impossible due to the non-zero sine terms in Eqs.~(\ref{eq:coefficients:v},\ref{eq:coefficients:t}), and furthermore, an analytical approach cannot be taken with the projected model in this parameter region.
Therefore, we studied the eigenstate localization properties numerically using standard probes for localization.
The inverse participation ratio (IPR)~\cite{wegner1980inverse,castellani1986multifractal,evers2008anderson} for an eigenstate \(\psi_n\),
\begin{gather}
  \label{eq:ipr}
  \mathrm{IPR} = \sum_{n}\abs{\psi_{n}}^{4} \sim L^{-\tau},
\end{gather}
is commonly used to quantify the degree of localization of a wave function in a system of size \(L\).
The quantity \(\tau\) is a scaling exponent of the \(\mathrm{IPR}\).
In the thermodynamic limit, one can define \(\tau\) explicitly as follows~\cite{vicsek1992fractal},
\begin{gather}
  \label{eq:tauexplicit}
  \tau = \lim_{L\to\infty}\frac{1}{\ln(1/L)}\ln\operatorname{IPR}.
\end{gather}
We rescale the eigenspectrum to fit into the interval \([0, 1]\) and split into equidistant bins \(e\).
Then we calculate the average \(\tau\) in a single bin \(e\), denoted as \(\mbeet\) as a function of lattice size \(L\).
It follows:
\begin{gather}
  \label{eq:asymptotic}
  \lim_{L\to\infty}\mbeet =
  \begin{dcases}
    0\text{, all states localized in \(e\),} \\
    \tau_{0}\text{, at least some states critical in \(e\)},
  \end{dcases}
\end{gather}
where \(\tau_{0}\) is a finite value between \(0\) and \(1\).
Suppose all states in a bin are localized.
Then we expect the following equation, which resembles Eq.~\eqref{eq:tauexplicit}, to hold as the lattice size changes,
\begin{gather}
  \label{eq:linearmodel}
  \mbeet = \frac{\kappa(e)}{\ln(1/L)} + \mathrm{intercept}, \quad \mathrm{intercept} = 0.
\end{gather}

The average \(\tau\) over the entire spectrum gives \(\mbet\). 
An important quantity in such a statistical approach is the \(R^2\) or goodness-of-fitting~\cite{steel1960principles}.
\(R^2\) reflects how well the data are described with a given statistical model; \(R^2 \approx 1\) concludes that states in bin \(e\) are localized in the thermodynamic limit, 
while \(R^2 \leq 0\) indicates that the set of numerical data does not follow the linear model in Eq.~\eqref{eq:linearmodel} at all.
Hence, in the thermodynamic limit, there are extended or critical states for which a different fitting model should be used.

We scanned the control parameter region given by \(0<\beta<\pi\) and \(0 < \lambda_2/\lambda_1 \leq 1\), discretized into \(25\times 25\) points.
Then the fitting procedure of \(\mbet\) is computed via Eq.~(\ref{eq:linearmodel}) for lattice sizes \(L = 2000, 3000, \dots 9000\) in steps of \(1000\).
The numerical results are summarized in Fig.~\ref{fig:phasediagram}, where it can be seen that all the eigenstates of the projected model in Eq.~\eqref{eq:pm} are localized in the thermodynamic limit away from the border of the phase diagram.

\subsection{Subdiffusive wavepacket spreading}

To further quantify the transport properties, we analyze the spreading of an initially localized wavepacket for weak quasiperiodic perturbation.
We use an initial state localized on a single site in the center of the lattice.
For convenience, we assign the lattice center to be the zero coordinate.
To quantify the spreading of the initial state, we compute the root-mean-square of the displacement \(\sigma(t)\):
\begin{gather}
  \label{eq:sigma-t}
  \sigma(t) = \left[\sum_{n\in\mathbb{Z}} (n - \langle n\rangle(t))^2 \abs{\psi_{n}(t)}^2 \right]^{1/2} \propto t^{\gamma},
\end{gather}
where \(\langle n\rangle\) is the average position defined as
\begin{gather}
  \langle n \rangle(t) = \sum_{n\in\mathbb{Z}} n \abs{\psi_n(t)}^2. \notag
\end{gather}

The displacement \(\sigma(t)\) provides the deviation of the particle's position from its average at time \(t\).
The absence of spreading indicates localization.
Also, the spreading always stops in finite systems once the boundaries are reached.
At an intermediate time, i.e., a period before the boundaries are reached, \(\sigma(t)\) is fitted by a power law with the exponent \(\gamma\) whose value indicates the type of transport: diffusive, subdiffusive, or ballistic~\cite{dominguez2019aubry}.

\begin{figure}
  \includegraphics[width = \linewidth]{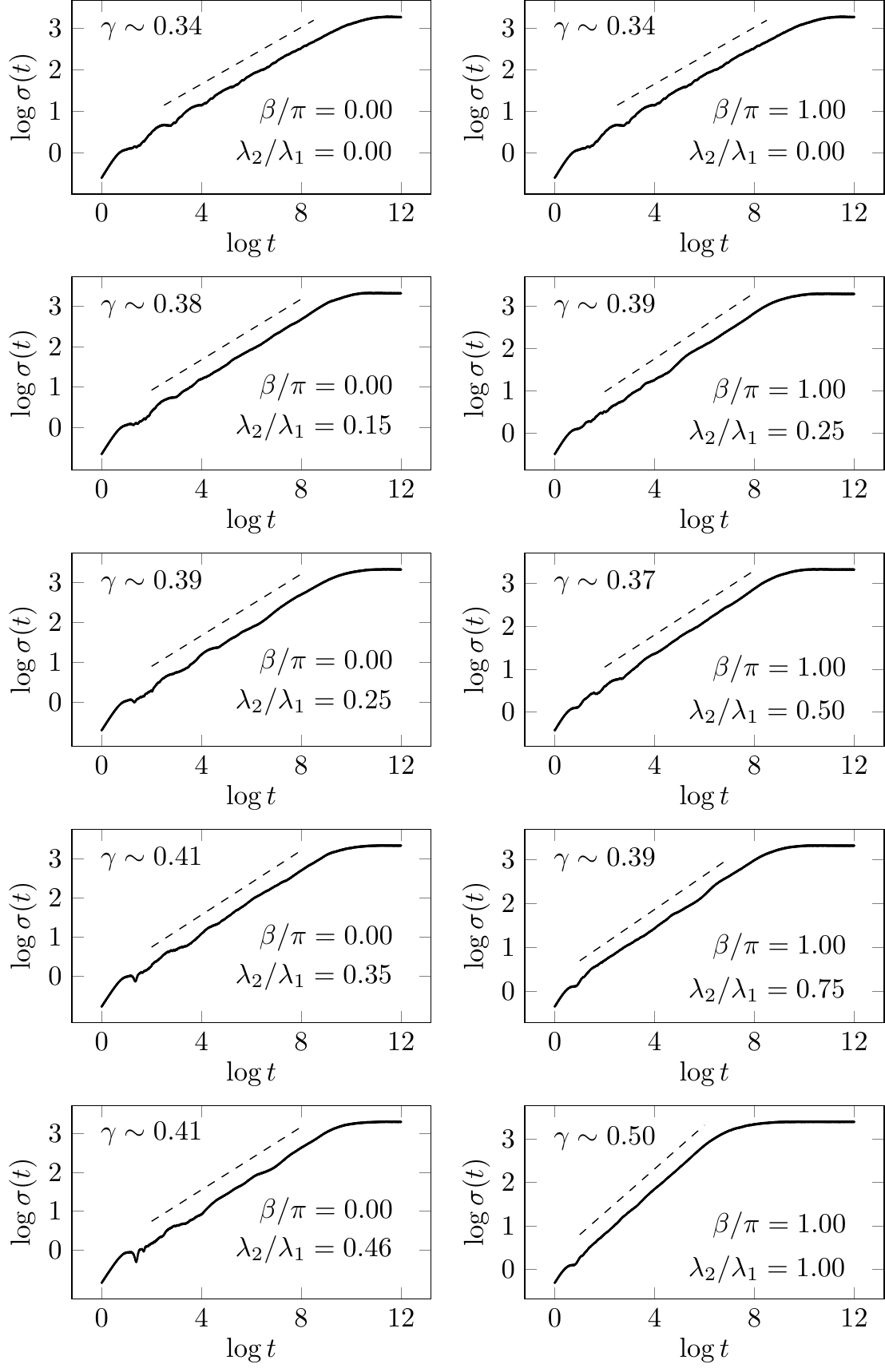}
  \caption{
    Log-log plots of the spreading of a wavepacket initially localized on a single site using \(\log_{10}\).
    The order of the plots follows the data in Table~\ref{table:subdiffusion}.
  }
  \label{fig:diffusion}
\end{figure}

\begin{figure}
  \includegraphics[width = \linewidth]{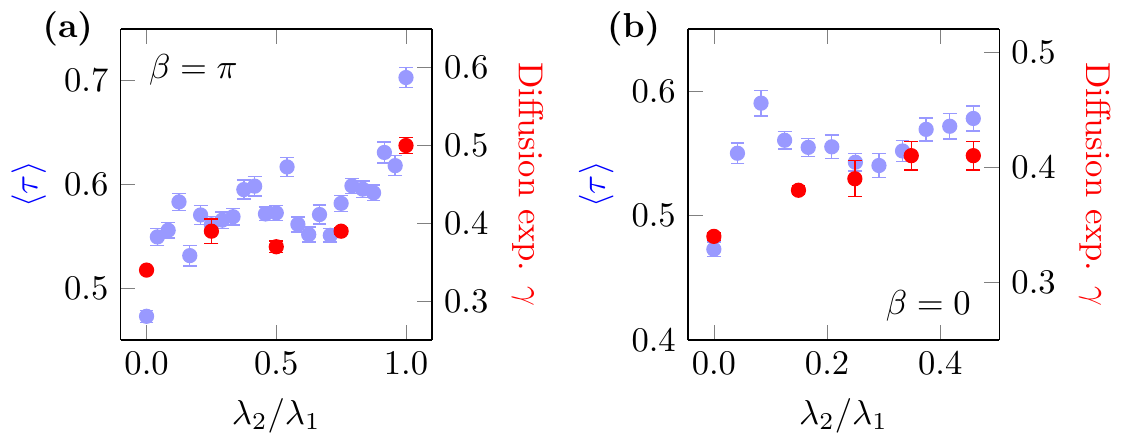}
  \caption{
    Positive correlation between the exponent \(\gamma\) in Eq.~\eqref{eq:sigma-t} (red points) and the exponent \(\tau\) averaged over the spectrum, \(\mbet\), in Eq.~\eqref{eq:linearmodel} (blue points).
    The values of \(\gamma\) are taken from Table~\ref{table:subdiffusion}.
    (a) \(\beta=\pi\).
    (b) \(\beta=0\).
  }
  \label{fig:taugamma}
\end{figure}

We have picked several points \((\lambda_2/\lambda_1, \beta/\pi)\) corresponding to the extended Harper model mapping and featuring critical eigenstates at the border of the phase diagram (for localized cases, the spreading stops as discussed above).
The results of the fitting are provided in Table~\ref{table:subdiffusion} and the details of the wavepacket spreading are shown in Fig.~\ref{fig:diffusion} for system size \(L = 12801\).
In all cases, we see a clear subdiffusion, whose exponent \(\gamma\) depends on the position on the border of the phase diagram, i.e., the values of \(\lambda_2/\lambda_1, \beta/\pi\).
For \(\lambda_2/\lambda_1=0\), \(\beta\) drops out of the Hamiltonian in Eq.~\eqref{eq:pm}, and consequently, the diffusion exponent \(\gamma\) does not depend on \(\beta\).
Furthermore, we observe a positive correlation between \(\gamma\) and \(\mbet\) in Fig.~\ref{fig:taugamma}.
This suggests that the transport properties are strongly affected by the details of the profiles of the critical eigenstates.

\begin{table}
  \centering
  \setlength{\tabcolsep}{5pt}
  \renewcommand{\arraystretch}{1.5}
  \begin{tabular}{c|c||c|c}
    \noalign{\smallskip}\noalign{\smallskip}\hline\hline
    \((\lambda_{2}/\lambda_{1}, \beta/\pi)\) & \(\gamma \pm \Delta\gamma\) & \((\lambda_{2}/\lambda_{1}, \beta/\pi)\) & \(\gamma \pm \Delta\gamma\) \\
    \hline
    \((0.00, 0.00)\) & \(0.34\pm 0.01\) & \((0.00, 1.00)\) & \(0.34\pm 0.01\) \\
    \hline
    \((0.15, 0.00)\) & \(0.38\pm 0.01\) & \((0.25, 1.00)\) & \(0.39\pm 0.04\) \\
    \hline
    \((0.25, 0.00)\) & \(0.39\pm 0.04\) & \((0.50, 1.00)\) & \(0.37\pm 0.01\) \\
    \hline
    \((0.35, 0.00)\) & \(0.41\pm 0.03\) & \((0.75, 1.00)\) & \(0.39\pm 0.02\) \\
    \hline
    \((0.46, 0.00)\) & \(0.41\pm 0.03\) & \((1.00, 1.00)\) & \(0.50\pm 0.01\) \\
    \hline\hline
  \end{tabular}
  \caption{
    Diffusion exponent \(\gamma\) in Eq.~\eqref{eq:sigma-t} for various points on the border of the phase diagram in Fig.~\ref{fig:phasediagram}.
    Except for \(\lambda_{2}/\lambda_{1} = 1\) and \(\beta = \pi\), all results show that the critical states support clear subdiffusive transport.
    For the off-diagonal Harper model, \(\beta=\pi,\lambda_1=\lambda_2\), the diffusion exponent is either diffusive or subdiffusive but very close to diffusive, 
    similar to the case of the Aubry--Andr\'e--Harper model at the critical point~\cite{hiramoto1988dynamics,wilkinson1994spectral}.
  }
  \label{table:subdiffusion}
\end{table}

\section{Finite perturbation}
\label{sec:finite}

In the limit of weak interaction, we saw the emergence of entirely critical spectra for specific values of the relative potential strength \(\lambda_2/\lambda_1\) and the phase difference \(\beta\).
As we now increase the perturbation strength and make it finite, we expect the system to localize for large enough values of \(\lambda_{1,2}\).
The open question is what happens to the critical states for moderate values of \(\lambda_{1,2}\), and also whether any extended states might emerge.
Also, we do not expect any change in the localization properties of the states that were already localized for weak quasiperiodic perturbation.

For finite potential strengths, the projected model description is not valid anymore and thus we have to focus on the original Hamiltonian, \(\mh = \mhabf + W\).
The Hamiltonian \(\mh\) can be expressed as a non-Abelian Aubry--Andr\'e--Harper model~\cite{wang2016phase} with a quasiperiodic diagonal block and a fixed hopping block:
\begin{gather}
  E\psi_{n} = A_{n}\psi_{n} + B\psi_{n+1} + B^{\dagger}\psi_{n-1},
\end{gather}
where \(\psi_n = (p_n, f_n)^T\), and \(A_{n}\) and \(B\) are (\(\Delta\equiv|\varepsilon_{a} - \varepsilon_{b}|\)):
\begin{gather}
  \label{eq:Exact}
  A_{n} = \frac{4}{\Delta}
  \begin{bmatrix}
    W_{1}(n) & 0 \\
    0 & W_{2}(n)
  \end{bmatrix},
  \quad B =
  \begin{bmatrix}
    1 & -1 \\
    1 & -1
  \end{bmatrix}.
\end{gather}
We focus on the parameter regions where the projected model [Eq.~\eqref{eq:pm}] for weak disorder hosts critical states.
We expect that the localized eigenstates for weak disorder remain localized for a finite potential, as we have checked below for several points in the parameter space.
It turns out to be convenient to work with the \emph{semi-detangled} Hamiltonian~\cite{danieli2021nonlinear}, which is defined by reverting the second local unitary \(U_{2}\) defined in Sec.~\ref{sec:model}.
This defines the semi-detangled basis \(\{u_{n}, d_{n}\}\) and gives the new, semi-detangled Hamiltonian,
\begin{gather}
  \label{eq:ham-sd}
  \mhsd = U_{1} \mhfd U_{1}^{\dagger} + U_{2}^{\dagger}WU_{2}.
\end{gather}
The semi-detangled wavefunction amplitudes \(u_{n}\) and \(d_{n}\) are related to \(\{p_{n}, f_{n}\}\) as follows (for \(\theta=\pi/4\)):
\begin{gather}
  u_{n} = \frac{1}{\sqrt{2}}(p_{n} - f_{n}) \quad\text{and}\quad d_{n} = \frac{1}{\sqrt{2}}(p_{n} + f_{n}).
\end{gather}
For convenience, let \(\varepsilon = (\varepsilon_{b} + \varepsilon_{a})/2\) and \(t = (\varepsilon_{b} - \varepsilon_{a})/2\).
Below we discuss several parameter values of \(\beta,\lambda_2,\lambda_1\) that allow for some analytical results.

\subsubsection{\(\beta = \pi\) and \(\lambda_{1} = \lambda_{2} = \lambda\)
\label{sec:claim}}

For \(\beta = \pi\) and \(\lambda_{1} = \lambda_{2} = \lambda\), we recall that the projected model is critical and maps onto the off-diagonal Harper model, which supports the strongest hopping among other parameters with vanishing onsite potential.
Now we move on to the finite perturbation.
The semi-detangled Hamiltonian in Eq.~\eqref{eq:ham-sd} takes the following form:
\begin{align}
  \left[E - \varepsilon\right] u_{n} &= \lambda\cos(2\pi\alpha n)d_{n} + td_{n+1},\\
  \left[E - \varepsilon\right] d_{n} &= \lambda\cos(2\pi\alpha n)u_{n} + tu_{n-1},
\end{align}
where \(\lambda_{1,2} = \lambda\).
We can eliminate one of the amplitudes \(u_n\) or \(d_n\) by substituting one of the equations into the other. 
If we choose to eliminate \(u_n\), we get the following effective equation:
\begin{align}
  \label{eq:EHMlike}
  \tilde{E} d_{n} &= \cos(4\pi\alpha n)d_{n} \vphantom{\frac{2t}{\lambda}} \\
  & +\frac{2t}{\lambda}\left[ \cos(2\pi\alpha (n-1))d_{n-1} + \cos(2\pi\alpha n)d_{n+1} \right], \notag
\end{align}
where the eigenvalue \(\tilde{E}\) is
\begin{gather}
  \tilde{E} = \frac{2}{\lambda^2}\left[(E - \varepsilon)^2 - t^2 - \frac{\lambda^{2}}{2}\right].
\end{gather}
This eigenequation looks very similar to the extended Harper model, Eq.~\eqref{eq:EHM}~\cite{avila2017spectral}, except for the spatial frequency of the onsite potential that is double that of the hopping.
Also, the potential is twice as large as in the extended Harper model.

The non-Abelian Aubry--Andr\'e--Harper model for this case has been studied numerically~\cite{degottardi2013majorana, wang2016phase}.
The critical eigenstates show up over the entire spectrum until \(\lambda\) reaches the value of the flatband bandgap \(\Delta=|\varepsilon_{b} - \varepsilon_{a}|\).
Then the CIT occurs once \(\lambda\) is equal to \(\Delta=|\varepsilon_{b} - \varepsilon_{a}|\) for the entire spectrum.
We remark that this result coincides with the extended Harper model estimate of the transition, \(2t/\lambda=1\), if we neglect the differences between our model, Eq.~\eqref{eq:EHMlike}, and the true extended Harper model, Eq.~\eqref{eq:EHM}.

\subsubsection{\(\beta = 0\) and \(\lambda = \lambda_{1} = \lambda_{2}\)
\label{sec:clspreserve}}

For \(\beta = 0\) and \(\lambda = \lambda_{1} = \lambda_{2}\), we obtain the following Hamiltonian in the semi-detangled basis \(u_n, d_n\):
\begin{align}
  \tilde{E}u_{n} &= \lambda\cos(2\pi\alpha n )u_{n} + t d_{n+1},\\
  \tilde{E}d_{n} &= \lambda\cos(2\pi\alpha n )d_{n} + t u_{n-1},
\end{align}
where \(\tilde{E} = E - \varepsilon\) and \(\lambda_{1,2} = \lambda\).
Again, we can eliminate one of the amplitudes through substitution.
Choosing to eliminate \(u_n\) leads to a diagonal problem for \(d_n\):
\begin{equation}
  \tilde{E}d_{n} = \left[\lambda\cos(2\pi\alpha n) + \frac{t}{\tilde{E} - \lambda\cos(2\pi\alpha(n-1))}\right]d_{n},
\end{equation}
and a compact localization of the eigenstates for any \(\lambda\).
We recall that we have already found compact localization for infinitesimal parameter values (see Sec.~\ref{sec:b0-l12}).
This compact localization survives for any potential strength.

\subsubsection{Emergence of fractality edges}

We now consider all the other values of \(\beta, \lambda_{1,2}\) where we observed critical states for weak perturbation.
In the semi-detangled basis \(u_n, d_n\), we obtain the following Hamiltonian:
\begin{align*}
  \tilde{E}u_{n} &= \frac{W_{1}(n) + W_{2}(n)}{2}u_{n} + \frac{W_{1}(n) - W_{2}(n)}{2}d_{n} + td_{n+1}, \\
  \tilde{E}d_{n} &= \frac{W_{1}(n) + W_{2}(n)}{2}d_{n} + \frac{W_{1}(n) - W_{2}(n)}{2}u_{n} + tu_{n-1}.
\end{align*}
\(\tilde{E}\) is \(E - \varepsilon\) and \(W_{1,2}\) are the quasiperiodic fields given in Eq.~\eqref{eq:fields}.
We see that the above effective equations are a mixture of the one given in Sec.~\ref{sec:claim} and Sec.~\ref{sec:clspreserve}.
This implies that the behavior of an eigenstate is at most critical.

To look into the properties of the states, we have to resort to numerical analysis on the full model in Eq.~\eqref{eq:Exact} to analyze their localization properties.
Flatband energies are fixed to \(\varepsilon_{a} = -1\) and \(\varepsilon_{b} = 2\) with bandgap \(\Delta=3\).
We consider several points of \(\beta, \lambda_{1,2}\) to probe the different regions of the phase diagram with critical states and repeat the computation of the IPR of the eigenstates.
Then we apply the linear model introduced in Eq.~\eqref{eq:linearmodel}, and introduce \(50\) energy bins for the eigenenergy range rescaled to fit between \(0\) and \(1\), which is referred to as the rescaled energy.
This allows us to distinguish localized and critical states as a function of eigenenergy.
The lattice sizes (in unit cells) are \(L = 2584, 3283, 4181, 4832, 5473\), and \(6765\).
Since the exact model has two sites for each unit cell, the size of the Hamiltonian matrix is \(2L\times 2L\).

The generic observation is that the critical spectrum of the projected model is replaced with a mixed one, comprising partially critical and partially localized states depending on the eigenenergy.
We dub the border between critical and localized states as \emph{fractality edges} by analogy with mobility edges separating localized and extended states.

\begin{figure}
  \includegraphics[width = \linewidth]{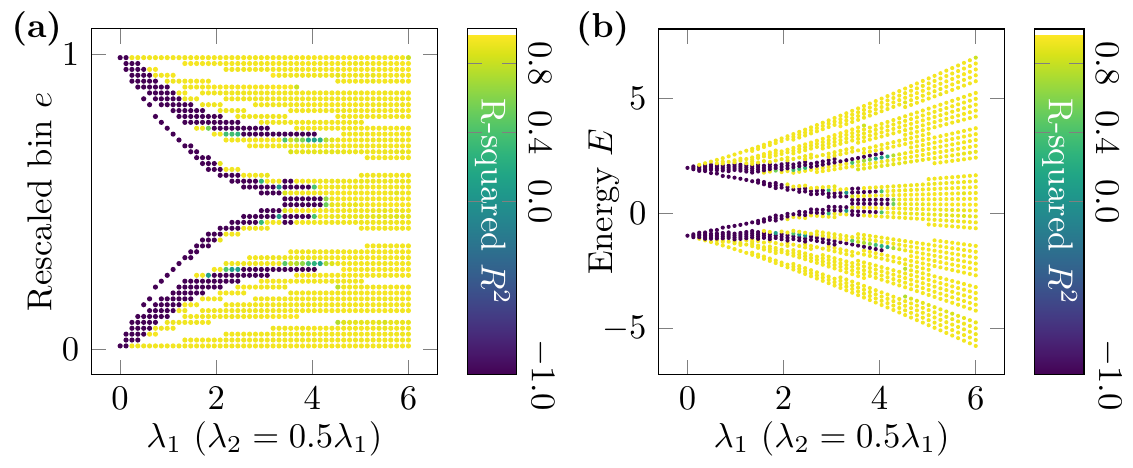}
  \caption{
    (a) Fractality edges in the exact model at \(\beta/\pi = 1, \lambda_{2}/\lambda_{1} =0.5\).
    Every \(R^2\) lower than or equal to zero is revalued to \(-1\) for the purpose of clear distinction between the localized and critical regions in the figure. 
    The \(-1\) \(R^2\) value implies that states are critical.
    For \(R^2 \approx 1\), the states are localized.
    (b) Fractality edges shown for the original, non-rescaled energy spectrum.
  }
  \label{fig:BetaPI0P50}
\end{figure}

First, let \(\beta = \pi\) and \(\lambda_{2}/\lambda_{1} = 0.5\).
Figure~\ref{fig:BetaPI0P50} shows the emergence of fractality edges for a finite strength of perturbation \(\lambda_{1,2}\).
We note that the critical states extend to values \(\lambda_1 \geq \Delta\), at variance with the case in Sec.~\ref{sec:claim} where all states localize for \(\lambda_1 > \Delta\).

\begin{figure}
  \includegraphics[width = \linewidth]{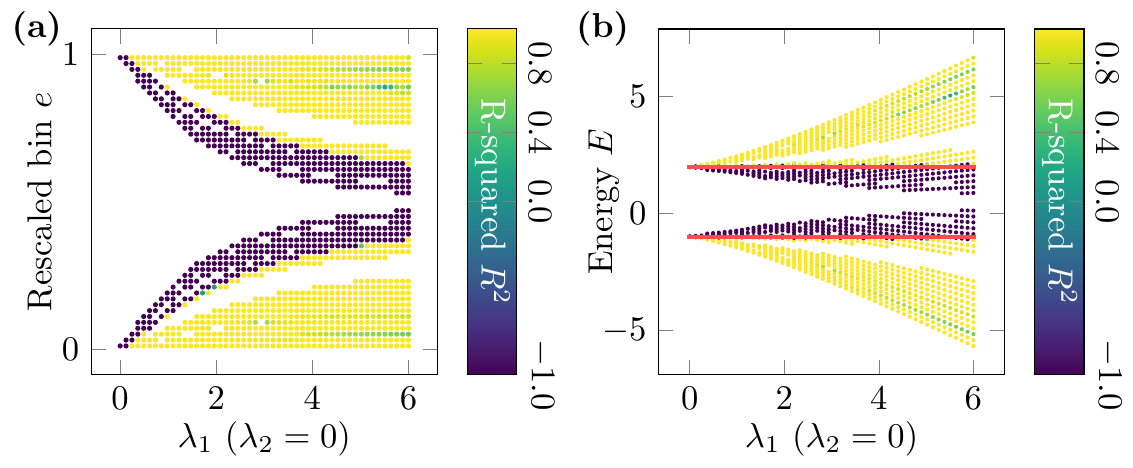}
  \caption{
    (a) Fractality edges in the exact model at \(\beta/\pi = 1\) and \(\lambda_{2}/\lambda_{1} = 0\).
    The value of \(\beta\) is irrelevant, as discussed in the text.
    (b) Fractality edges shown in the original, non-rescaled energy spectrum.
    All eigenstates between the flatband energies are always critical, while the eigenstates outside are localized.
    The red lines are flatband energies \(\varepsilon_{a} = -1\) and \(\varepsilon_{b} = 2\).
  }
  \label{fig:BetaPI0P00}
\end{figure}

Now let \(\lambda_{2} = 0\), i.e., one of the quasiperiodic fields is turned off completely, and then similarly to the case of the projected model, the phase difference \(\beta\) is irrelevant.
This can be seen from Eq.~\eqref{eq:Exact}, where the Hamiltonian remains unchanged as \(\beta\) is completely removed due to the absence of \(\lambda_{2}\).
As is clear from Fig.~\ref{fig:BetaPI0P00}, the fractality edges are present for any potential strength \(\lambda_1\) and coincide with the flatband energies;
as a result, the critical states are always located between the original flatbands.

For \(\beta = 0\), we observe fractality edges as well for \(\lambda_{2}/\lambda_{1}\leq (\lambda_{2}/\lambda_{1})_{c} \approx 0.468\) (not shown).
For larger ratios, no fractality edges emerge, and all the states are localized.

\section{Conclusions}
\label{sec:conclusion}

We have investigated the effect of quasiperiodic perturbation on all-bands-flat systems using a two-leg ladder with both bands flat as an example.
Choosing the local unitary transformation with the angles that maximize the hopping in the effective Hamiltonian, we identified parameter regions supporting critical states with subdiffusive transport for weak quasiperiodic perturbation.
The emergence of critical states in 1D systems is quite different from both the random disorder case in ABF networks~\cite{cadez2021metalinsulator}, which only allows for localized states, and the conventional Aubry--Andr\'e model, which features a metal--to--insulator transition.
The critical states were found when the phase difference \(\beta=0, \pi\) or the amplitude ratio \(\lambda_2/\lambda_1\) is zero.
Additionally, for \(\beta=0\), a phase transition between localized and critical states, or CIT, was observed.
The transition point \((\lambda_2/\lambda_1)_c\) depends on the irrational spatial frequency \(\alpha\).
Away from these parameter values, all the states were localized in the thermodynamic limit, which we have shown numerically.
For finite potential strength, we have discovered fractality edges, an energy dependent CIT---that disappear in most cases as the strength of the potential is increased.
We have also identified a case where the critical states persist for arbitrary quasiperiodic potential strengths.

Understanding the origin of this effect is an open question worth investigating.
Our results generically align with the suggestion in Ref.~\onlinecite{nosov2019correlation} that critical states can be induced by adding correlations to disorder.
It might also be interesting to reveal a connection, if any, to the Rosenzweig--Porter model, which also displays critical states~\cite{kravtsov2015random,khaymovich2020fragile}.
Adding interactions and considering the case of many particles might also be of interest.

While preparing this manuscript, we became aware of the work~\cite{sharma2022flat} that reached similar conclusions for a diamond chain based ABF model in the presence of quasiperiodic perturbation.
We provide some basic analysis of this model using our framework in Appendix~\ref{app:diamond}.

\begin{acknowledgments}
  This work was supported by the Institute for Basic Science, Project Code (Project No.~IBS-R024-D1).
\end{acknowledgments}

\appendix

\section{Generic \(\mathrm{SU}(2)\) local unitary transformation}
\label{app:SU2}

In the main text, the local unitary transformations \(U(\theta)\) used to construct ABF Hamiltonians are taken to be real, i.e., as an element of the \(\mathrm{SO}(2)\) group, parameterized by angle \(\theta\).
However, the most generic possible form of \(U\) producing ABF networks is an element of \(\mathrm{SU}(2)\), parameterized by angle \(\theta\) and phases \(\eta,\zeta\):
\begin{gather}
  U(\theta) = 
  \begin{bmatrix}
    e^{i\zeta}\cos(\theta) & e^{-i\eta}\sin(\theta) \\
    -e^{i\eta}\sin(\theta) & e^{-i\zeta}\cos(\theta)
  \end{bmatrix}.
\end{gather}
Using this general transformation to derive the projected model gives the hopping term \(t_{n}\) in Eq.~\ref{eq:pm} with an extra complex phase \(t_{n}\exp\left(-i\Theta\right)\), 
where \(\Theta = \zeta_{2} - \zeta_{1} - \eta_{2} + \eta_{1}\), and subscripts \(1\) and \(2\) indicate the first and second local unitary matrices \(U_{1,2}\).
The phase \(\exp\left(-i\Theta\right)\) can then be eliminated with an appropriate local unitary transformation in the form of Eq.~\ref{eq:eliminatephase} (see Appendix~\ref{app:self-dual}), after which the effective model reduces to the case of \(\mathrm{SO}(2)\) transformations.

\section{Full expressions for \(v_{s,c},t_{s,c}\)}
\label{app:vt-expr}

The onsite and hopping terms of the projected model for generic local unitary transformation angles \(\theta_{1,2}\) are as follows:
\begin{align}
  v_{n} &= \cos^{2}\theta_{1}\cos^{2}\theta_{2}\cos(2\pi\alpha n) \\
        &+ \frac{\lambda_{2}}{\lambda_{1}}\cos^{2}\theta_{1}\sin^{2}\theta_{2}\cos(2\pi\alpha n + \beta) \notag \\
        &+ \sin^{2}\theta_{1}\sin^{2}\theta_{2}\cos(2\pi\alpha (n-1)) \notag \\
        &+ \frac{\lambda_{2}}{\lambda_{1}}\sin^{2}\theta_{1}\cos^{2}\theta_{2}\cos(2\pi\alpha(n-1) + \beta),\notag \\
  t_{n} &= \frac{1}{4}\sin 2\theta_{1}\sin 2\theta_{2}\cos(2\pi\alpha n) \\
        &- \frac{\lambda_{2}}{4\lambda_{1}}\sin 2\theta_{1}\sin 2\theta_{2}\cos(2\pi\alpha n + \beta). \notag
\end{align}
Using the identities
\begin{align*}
  \cos(x + y) &= \cos x \cos y - \sin x \sin y,\\
  \sin(x + y) &= \sin x \cos y - \cos x \sin y,
\end{align*}
we obtain the coefficients \(v_{s,c}\) and \(t_{s,c}\) in Eqs.~(\ref{eq:coefficients:v},\ref{eq:coefficients:t}) as follows:
\begin{align}
  v_{s} &= \sin(\pi\alpha)\left(\sin^2\theta_{1}\sin^2\theta_{2} - \cos^2\theta_{1}\cos^2\theta_{2}\right) \vphantom{\frac{\lambda_{2}}{\lambda_{1}}} \\
        &+ \frac{\lambda_{2}\cos\beta}{\lambda_{1}}\sin(\pi\alpha)\left(\sin^2\theta_{1}\cos^2\theta_{2} - \cos^2\theta_{1}\sin^2\theta_{2}\right) \notag \\
        &- \frac{\lambda_{2}\sin\beta}{\lambda_{1}}\cos(\pi\alpha)\left(\cos^2\theta_{1}\sin^2\theta_{2} + \sin^2\theta_{1}\cos^2\theta_{2}\right) \notag \\
  v_{c} &= \cos(\pi\alpha)\left(\cos^2\theta_{1}\cos^2\theta_{2} + \sin^2\theta_{1}\sin^2\theta_{2}\right) \vphantom{\frac{\lambda_{2}}{\lambda_{1}}} \\
        &+ \frac{\lambda_{2}\sin\beta}{\lambda_{1}}\sin(\pi\alpha)\left(\sin^2\theta_{1}\cos^2\theta_{2} - \cos^2\theta_{1}\sin^2\theta_{2}\right) \notag \\
        &+ \frac{\lambda_{2}\cos\beta}{\lambda_{1}}\cos(\pi\alpha)\left(\cos^2\theta_{1}\sin^2\theta_{2} + \sin^2\theta_{1}\cos^2\theta_{2}\right) \notag \\
  t_{s} &= \frac{1}{4}\sin 2\theta_{1}\sin 2\theta_{2}\frac{\lambda_{2}\sin\beta}{\lambda_{1}} \\
  t_{c} &= \frac{1}{4}\sin 2\theta_{1}\sin 2\theta_{2}\left(1 - \frac{\lambda_{2}\cos\beta}{\lambda_{1}}\right).
\end{align}

\section{Self-duality of the off-diagonal Harper model}
\label{app:self-dual}

For \(\beta = \pi\) and \(\lambda_{1} = \lambda_{2}\), the projected model in Eq.~\eqref{eq:pm} has a zero onsite potential \(v_n\equiv 0\) and quasiperiodic hopping \(t_{n} = \cos(2\pi\alpha n)/2\):
\begin{gather}
  \label{eq:hm-off-d}
  E a_{n} = t_{n}a_{n+1} + t_{n-1}a_{n-1}.
\end{gather}
This model, known as the off-diagonal Harper model~\cite{han1994critical,kraus2012topological}, is self-dual under a transformation similar to that of the Aubry--Andr\'e model.
We provide here the details of this transformation.
The starting point is a conventional Fourier transform:
\begin{gather}
  \ket{k} = \frac{1}{\sqrt{L}}\sum_{n}e^{i2\pi\alpha n k}\ket{a_{n}}.
\end{gather}
Applying this to the above model in Eq.~\eqref{eq:hm-off-d}, we find
\begin{align}
  H &= \frac{1}{2}\sum_{n} \cos(2\pi\alpha n) \left[\ketbra{a_{n}}{a_{n+1}} \vphantom{\frac{1}{2}}{+} \ketbra{a_{n+1}}{a_{n}}\right] \\
    &= \frac{1}{2}\sum_{k} e^{-i\pi\alpha} \ketbra{k+1}{k} \cos(\pi\alpha(2k+1)) \notag \\
    &+ e^{i\pi\alpha} \ketbra{k}{k+1} \cos(\pi\alpha(2k+1))\label{eq:transformation}. 
\end{align}
Eq.~\eqref{eq:transformation} is almost identical to Eq.~\eqref{eq:hm-off-d} except for the additional phase factors \(e^{\pm i\pi\alpha}\).
Then we apply the local unitary transformation \(R\) to remove the redundant phase factors:
\begin{gather}
  \label{eq:eliminatephase}
  R = \sum_{k} \ketbra{k+1}{k} e^{-i\pi\alpha/2} + \ketbra{k}{k+1} e^{i\pi\alpha/2}.
\end{gather}
This gives us the equivalent off-diagonal Harper model that we started with, as:
\begin{gather*}
  H^{\prime} = \frac{1}{2}\sum_{k}\cos(2\pi\alpha k  + \pi\alpha)\left(\ketbra{k+1}{k} + \ketbra{k}{k+1} \right).
\end{gather*}

\section{Diamond chain ABF: Fractality edges and multifractal states at \(E=0\)}
\label{app:diamond}

The interesting point the authors of Ref.~\cite{sharma2022flat} have claimed is the existence of multifractal states in the case of an antisymmetric quasiperiodic perturbation (see the reference for details).
The authors numerically tested the model and observed the multifractal states at around \(E = 0\) and \(\lambda \lessapprox 2\).
We consider their model theoretically to explain their numerical results.

For non-zero energy cases, the number of variables can be reduced via substitution in making the effective model.
The recurrence relations of the asymmetric quasiperturbed model are given as follows:
\begin{align}
  Eu_{n} &= -(c_{n} + c_{n-1}) + \lambda\cos(2\pi\alpha n) u_{n}, \\
  Ec_{n} &= -(u_{n+1} - d_{n+1} + u_{n} + d_{n}), \\
  Ed_{n} &= -(c_{n} - c_{n-1}) - \lambda\cos(2\pi\alpha n) d_{n}.
\end{align}
Let us introduce the following local unitary transformation as in Ref.~\onlinecite{flach2014detangling} and remove \(c_{n}\) via substitution:
\begin{gather*}
  p_{n} = \frac{1}{\sqrt{2}}(u_{n} + d_{n})\\
  f_{n} = \frac{1}{\sqrt{2}}(u_{n} - d_{n}).
\end{gather*}
Then we get recurrence relations consisting of \(p_{n}\) and \(f_{n}\) sites:
\begin{align}
  Ep_{n} &= \frac{2}{E}(p_{n} + f_{n+1}) + \lambda\cos(2\pi\alpha n)f_{n}, \\
  Ef_{n} &= \frac{2}{E}(p_{n-1} + f_{n}) + \lambda\cos(2\pi\alpha n)p_{n}.
\end{align}
Making another substitution, we clear out one of the variables and get the effective model composed of a single site variable:
\begin{align}
  \tilde{E} p_{n} &= \cos(4\pi\alpha n)p_{n} \\
  &+  \frac{4}{\lambda E}\left[\cos(2\pi\alpha (n+1))p_{n+1} + \cos(2\pi\alpha n)p_{n-1}\right], \notag
\end{align}
where the eigenvalue \(\tilde{E}\) is:
\begin{gather}
  \tilde{E} = \frac{2}{\lambda^{2}}\left[\left(E - \frac{2}{E}\right)^2 - \frac{4}{E^2} - \frac{\lambda^2}{2}\right].
\end{gather}
This model is equivalent to Eq.~\eqref{eq:EHMlike}.
Based on the claim we made in Section~\ref{sec:claim}, the CIT appears at \(4/(|E|\lambda_{c}) = 1\).
However, \(\lambda_{c}\) is energy dependent, which consequently gives the fractality edges.
The explicit equation of the fractality edges seem to match the numerical results of Ref.~\cite{sharma2022flat} (not shown).

On the other hand, zero energy works differently.
Setting \(E = 0\) gives different recurrence relations:
\begin{align}
  c_{n-1} + c_{n} &= \lambda\cos(2\pi\alpha n)u_{n}, \\
  0 &= -(u_{n} + d_{n} + u_{n+1} - d_{n+1}), \\
  c_{n-1} - c_{n} &= \lambda\cos(2\pi\alpha n)d_{n}.
\end{align}
After the same local unitary transformation as above, we get the following:
\begin{align}
  \sqrt{2}c_{n-1} &= \lambda\cos(2\pi\alpha n)p_{n}, \\
  p_{n} &= f_{n+1}, \\
  \sqrt{2}c_{n} &= \lambda\cos(2\pi\alpha n)f_{n}.
\end{align}
Again through a substitution, we get the effective model which is equivalent to the off-diagonal Harper model having self-duality as shown in Appendix~\ref{app:self-dual}:
\begin{gather}
  0 = \lambda\cos(2\pi\alpha n)p_{n-1} + \lambda\cos(2\pi\alpha(n+1))p_{n+1}.
\end{gather}

\bibliography{general,flatband,frustration,mbl}

\end{document}